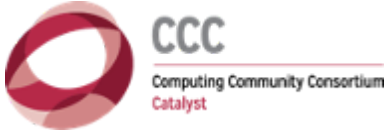

# Infrastructure for Artificial Intelligence, Quantum and High Performance Computing

*A Computing Community Consortium (CCC) Quadrennial Paper*

*William Gropp (University of Illinois at Urbana-Champaign), Sujata Banerjee (VMware Research) and Ian Foster (University of Chicago)*

High Performance Computing (HPC), Artificial Intelligence (AI)/Machine Learning (ML), and Quantum Computing (QC) and communications offer immense opportunities for innovation and impact on society. Researchers in these areas depend on access to computing infrastructure, but these resources are in short supply and are typically siloed in support of their research communities, making it more difficult to pursue convergent and interdisciplinary research. Such research increasingly depends on complex workflows that require different resources for each stage. This paper argues that a more-holistic approach to computing infrastructure, one that recognizes both the convergence of some capabilities and the complementary capabilities from new computing approaches, be it commercial cloud to Quantum Computing, is needed to support computer science research.

The types of infrastructure needed to support HPC and AI/ML share many features; GPU systems originally developed for HPC have become essential for ML, and those systems have further been optimized for ML, with features now being applied to HPC simulations. Further, techniques from both areas are being combined to create new capabilities. Quantum Computing is further from application, but many expect it to add a complementary computing capability that will be used with HPC and other systems. While there are ongoing research and infrastructure initiatives in all three areas (e.g., AI institutes and the National Quantum Initiative Act), they are not coordinated with each other, making it harder to share resources and to combine HPC and AI (and later QC) to solve the most challenging problems.

HPC and AI/ML technologies have reached some level of maturity, but Quantum Computing and communication are still in the early stages of development. HPC is the most mature, even if continuing to evolve; AI/ML is in a mixed state, with the technology adequate to use in a wide range of applications, but with many important unanswered questions; for Quantum, there is not even agreement on device technology or clarity on computing models. Nevertheless, understanding how all three technologies might complement each other and share infrastructure requires early thinking and planning, especially as multiple federal and industry efforts are initiated to fund research and build new infrastructure and testbeds.

Given that HPC and AI/ML have existing synergies, we first address the needs of these communities. Both HPC and AI/ML have become mainstream and need to be broadly accessible to all researchers across the US. Both need to support the big missions (exascale simulations, training GPT3, etc.) and also the masses of people who need to use these capabilities on a daily basis. HPC, and especially the needs of simulation, are often the bigger driver of needs for massive hardware, while AI/ML is the bigger driver of needs for access to hardware at the low end. In addition, application software is often a bigger barrier than hardware, and is historically not supported well by the federal government. Now that HPC and AI are mainstream, a major challenge is to embed them into the workflows of every aspect of science. That will involve a combination of hardware and software.

We anticipate that the demand for HPC and AI/ML systems will continue to grow to address unmet needs. Additionally, the end of Dennard scaling is leading to "speciation" in computing. While there are some common elements (in HPC and AI), we are also seeing specialized solutions emerge (e.g., TPU and Cerebras for ML). The need is to support the success of computing (e.g., the broad and growing use of HPC in simulation), the emerging success in data science and in AI/ML, and future specialized accelerators, ranging from near-term to long term and highly speculative—e.g., QC, with neuromorphic somewhere in the middle. Just as the research community has devoted significant efforts to understanding the offload architectures and connectivity to existing and emerging accelerators, there needs to be new research to understand what this would look like for quantum accelerators of the future. While loose coupling may be the starting point, new offload architectures accounting for new workloads and new devices will need further investigation. Further, though quantum accelerators will not be suitable for all ML workloads, understanding where/if they might complement existing AI/ML infrastructure would be a good exploration to conduct.

**Recommendations**

Both computational simulation and AI have become vitally important, mainstream, general-purpose research tools. It is vital to US competitiveness that we ensure broad access to associated infrastructure, in the form of HPC, AI/ML and evolving quantum computing hardware. Industry also has a key role in deploying and making available such infrastructure, some of which is already happening.

Current access is inadequate, due to problems of capability (inadequate peak supply), capacity (inadequate aggregate supply), and usability (inadequate integration with tools and workflows used by different communities). Further, formal coordination between multiple agencies investing in disparate computing resources and testbeds is limited and needs to be fixed. In addition to funding each of the HPC/AI/ML/Quantum infrastructures separately, it will be important to provide funding to support collaborative work across these research communities, as well as, infrastructure support for using these in combined workflows.

It is essential that adequate cyberinfrastructure be provided to support computer science research and that this infrastructure exploit synergies between HPC and ML, while also providing for the integration of QC as that technology matures. Such cyberinfrastructure will be a combination of co-located resources at computing centers, coordinated access to focused testbeds and to commercial cloud resources. Only in this way will the needs of the most demanding, interdisciplinary research be met.

*This white paper is part of a series of papers compiled every four years by the CCC Council and members of the computing research community to inform policymakers, community members and the public on important research opportunities in areas of national priority. The topics chosen represent areas of pressing national need spanning various subdisciplines of the computing research field. The white papers attempt to portray a comprehensive picture of the computing research field detailing potential research directions, challenges and recommendations.*

*This material is based upon work supported by the National Science Foundation under Grant No. 1734706. Any opinions, findings, and conclusions or recommendations expressed in this material are those of the authors and do not necessarily reflect the views of the National Science Foundation.*

*For citation use: Banerjee S., Foster I., & Gropp W. (2020) Infrastructure for Artificial Intelligence, Quantum and High Performance Computing.*
*https://cra.org/ccc/resources/ccc-led-whitepapers/#2020-quadrennial-papers*